\documentclass[amsfonts,prl,aps,twocolumn]{revtex4}
\usepackage{amsmath}
\usepackage{amssymb}
\usepackage{graphicx}
\begin{document}
\title{Strong pressure-energy correlations in van der Waals liquids} 
\author{Ulf R. Pedersen, Nicholas P. Bailey, Thomas B. Schr{\o}der, and Jeppe C. Dyre}
\affiliation{DNRF Centre ``Glass and Time,'' IMFUFA, Department of Sciences, Roskilde University, Postbox 260, DK-4000 Roskilde,
Denmark} 
\date{\today}

\begin{abstract}
Strong correlations between equilibrium fluctuations of the configurational parts of pressure and energy are found in computer simulations of the Lennard-Jones liquid and other simple liquids, but not for hydrogen-bonding liquids like methanol and water. The correlations, that are present also in the crystal and glass phases, reflect an effective inverse power-law repulsive potential dominating fluctuations, even at zero and slightly negative pressure. In experimental data for supercritical Argon, the correlations are found to be approximately 96\%. Consequences for viscous liquid dynamics are discussed.
\end{abstract}


\maketitle

For any macroscopic system thermal fluctuations are small and apparently insignificant. That the latter is not the case was pointed out by Einstein, who showed that for any system in equilibrium with its surroundings, the specific heat is determined by the magnitude of the energy fluctuations. This result may be generalized, and it has long been well understood that linear-response quantities are determined by equilibrium fluctuations of suitable quantities \cite{lan70,han86,rei98}. One expects few new insights to come from studies of fluctuations in equilibrated systems. We here report strong correlations between instantaneous pressure and energy equilibrium fluctuations in one of the most studied models in the history of computer simulation, the Lennard-Jones liquid. These findings have significant consequences, in particular for the dynamics of highly viscous liquids.

\begin{figure}
\begin{center}
\includegraphics[width=8cm]{2007_correlations_fig_1a.eps} 
\includegraphics[width=9cm]{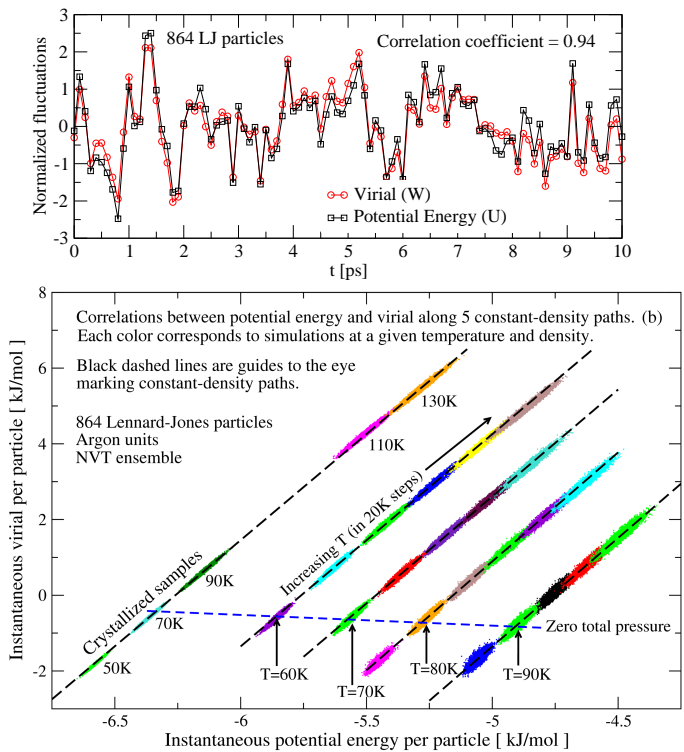}
\caption{Results from equilibrium molecular dynamics simulations of 864 particles interacting via the Lennard-Jones potential studied in the $NVT$ ensemble where ``Argon-units'' were used ($\sigma =0.34$nm, $\epsilon=0.997$kJ/mol). (a) Normalized fluctuations at $T=80K$ and zero average pressure (density = 34.6 mol/l) of the virial, W, $\Delta W(t)/\sqrt{\langle (\Delta W)^2 \rangle}$, and of the potential energy, $\Delta U(t)/\sqrt{\langle (\Delta U)^2 \rangle}$. The equilibrium fluctuations of virial and potential energy are strongly correlated, as quantified by the correlation coefficient: $R \equiv \langle \Delta W \Delta U\rangle/\sqrt{\langle (\Delta W)^2 \rangle\langle (\Delta U)^2 \rangle}=0.94$, where averages are over the full length of the simulation (10ns) after 10ns of equilibration. (b) Configurational parts of pressure and energy -- virial versus potential energy -- for several state points of the Lennard-Jones liquid. Each color represents simulations at one particular state point where each data point marks instantaneous values of virial and potential energy from a 10ns simulation. The black dashed lines mark constant density paths with the highest density to the upper left (densities: 39.8 mol/l, 37.4 mol/l, 36.0 mol/l, 34.6 mol/l, 32.6 mol/l). State points on the blue dashed line have zero average total pressure. The plot includes three crystallized samples (lower left corner).}
\label{figure_1}
\end{center}
\end{figure}

Using molecular dynamics \cite{all93,gromacs}, fluctuations were studied for $N=864$ particles interacting via the Lennard-Jones (LJ) pair potential \cite{len31} $\phi_{LJ}(r)=4\epsilon\,[(\sigma/r)^{12}-(\sigma/r)^6]$ in the NVT ensemble \cite{NoseHoover}, where $r$ is the distance between two particles. The configurational contribution to the instantaneous pressure defines the instantaneous virial $W(t)$ by \cite{all93} $p(t)V=Nk_BT(t)+W(t)$. Fig.~1(a) shows normalized instantaneous equilibrium fluctuations of $W(t)$ and the potential energy $U(t)$ for a simulation at zero average pressure. The two quantities correlate strongly. To study the correlations systematically, temperature was varied at five different densities. The results are summarized in Fig.~1(b), plotting instantaneous virial versus instantaneous potential energy, with each color representing equilibrium fluctuations at one particular temperature and density. The figure reveals strong $W,U$ correlations with correlation coefficients mostly above $0.9$, see table \ref{table:rgamma}. The results at a given density form approximate straight lines.  The data include slightly negative pressure conditions, as well as three instances of the crystallized liquid (lower left corner).

For any system with pair-wise interactions $W(t)=-\sum_{i<j}r_{ij}(t)\phi'(r_{ij}(t))/3$ \cite{han86,all93}. Perfect $W,U$ correlation applies if $\phi(r) = ar^{-n}+\phi_0$ in which case $\Delta W(t) =(n/3)\Delta U(t)$ where $\Delta W(t) \equiv W(t) - \langle W \rangle$, etc. An obvious first guess is therefore that the strong correlation directly reflects the $r^{-12}$ term in the LJ potential. That is not correct because the exponent $n=12$ implies a slope of $\gamma=4$ of the lines in Fig.~1(b); the observed slope is $\gamma=6$ ($\pm 10\%$,  see table \ref{table:rgamma}), corresponding to effective inverse power-law exponents $n\approx 18$. The repulsive core of the LJ potential ($r{}<2^{1/6}\sigma$), however, can be well approximated by $\phi_{pow}(r) = ar^{-n} + \phi_0$, with an exponent $n$ considerably \emph{larger} than 12 \cite{BenAmotz03,kan85}. If one requires that the 0'th, 1'st and 2'nd derivatives of the two potentials agree at $r=r_0$, one finds $n(r_0) = 6+12/[2-(r_0/\sigma)^6]$. Thus, $n(\sigma)=18$, whereas $n(0.969\sigma)=16.2$ (this is where $\phi_{LJ}=\epsilon$). 

To directly test whether the fluctuations are well described by an inverse power-law potential, we proceeded as follows.
A large number of configurations from the simulation of the zero-pressure state-point in Fig.~1(a) were stored. This time-series of configurations was analyzed by splitting the potential energy into two terms: $U(t) = U_{pow}(t) + U_{rest}(t)$, where  $U_{pow}(t)\equiv \sum_{i<j} \phi_{pow}(r_{ij}(t))$, i.e., the potential energy if the interatomic potential were an inverse power-law. Comparing $U_{pow}(t)$ and $U(t)$ it was found that the fluctuations were nearly identical, $\Delta U_{pow}(t)\approx \Delta U(t)$, with a correlation coefficient of 0.94. Applying the same procedure to the virial, we found $\Delta W_{pow}(t) \approx \Delta W(t)$ with a correlation coefficient of 0.99. These results prove that the repulsive core of the LJ potential dominates fluctuations, even at zero and slightly negative pressure, \emph{and} that at a given state point it is well-described by an inverse power-law potential. The fact that the repulsive forces
dominate the physics -- here the fluctuations -- confirms the philosophy of the well-known Weeks, Chandler, Andersen approximation \cite{wee71}.

It should be stressed that our approach is \emph{not} to choose a particular inverse power-law and analyze the results in terms of it. In fact, for an exact inverse power-law potential all data points in Fig. 1(b) would fall on the same line [$W(t)=\frac{n}{3}U(t)$]. Instead we simply study the equilibrium fluctuations at each state-point and find strong W,U correlations, which in turn can be explained by an effective inverse power-law dominating the \emph{fluctuations}. The effective inverse power law exponent is weakly state-point dependent, and the above explanation is consistent with the qualitative trends seen in table  \ref{table:rgamma}: Increasing temperature along an isochore or increasing density along an isotherm results in stronger correlation and smaller slopes, corresponding to a numerically smaller apparent exponent. This reflects particles approaching closer to each other, and thus $r_0$ decreasing ($n(r_0)$ decreasing) and the inverse power law being an even better approximation to the LJ potential close to $r_0$. We do find that $\gamma\rightarrow 4$ ($n\rightarrow 12$) at high temperatures and/or densities as expected, but only under quite extreme conditions, see table \ref{table:rgamma}. Along an isobar there is competition between the effects of density and temperature. Our results show that the density effect dominates: the correlation increases with \emph{decreasing} temperature. This, incidentally, is the limit of interest when studying highly viscous liquids (see below).

\protect{
\begin{table}
 \begin{tabular}{|c|c|c|c|c|}
  \hline
    $\rho = 34.6$ mol/l & 60K  & 80K & 100K & 1000K \\ 
  \hline
     R &  0.900 & 0.939 & 0.953 & 0.997  \\ 
  \hline
     $\gamma $ & 6.53 & 6.27 & 6.08 & 4.61 \\ 
  \hline
  \hline   
T =  130K & 32.6 mol/l  & 36.0 mol/l & 39.8 mol/l &  \\
  \hline
     R &  0.945 & 0.974 & 0.987 &  \\
  \hline
     $\gamma $ & 6.06 & 5.71 & 5.40 &  \\
  \hline   \hline
p = 0.0 GPa  & 60K  & 70K & 80K & 90K \\
  \hline
     R &  0.965 & 0.954 & 0.939 & 0.905  \\
  \hline
     $\gamma $ & 6.08 & 6.17 & 6.27 & 6.52 \\ 
  \hline
 \end{tabular}  
\caption{The correlation coefficient,  $R$, 
and the slope 
$\gamma \equiv \sqrt{\langle (\Delta W)^2 \rangle/\langle (\Delta U)^2 \rangle}$
along an isochore, an isotherm, and an isobar for the Lennard-Jones Liquid.
} \label{table:rgamma}
\end{table} 
}

\begin{figure}
\begin{center}
\includegraphics[width=8cm]{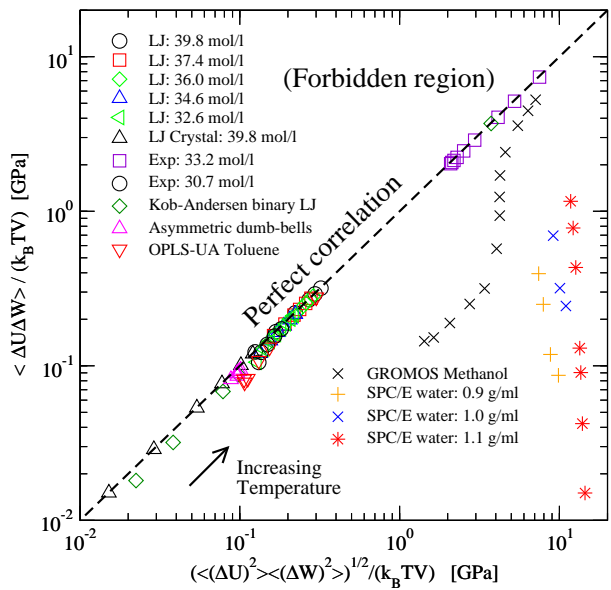}
\caption{$\langle \Delta W \Delta U \rangle/(k_BTV)$ plotted as a function of $( \langle (\Delta W)^2 \rangle\langle (\Delta U)^2 \rangle )^{1/2}/(k_BTV)$ for several liquids. If the correlation is perfect ($R=1$) the data fall on the diagonal. The region above the diagonal corresponds to $R>1$ and is thus forbidden. 
{\bf LJ:} Lennard-Jones results from the simulations reported in Fig. 1. Similar results were found in the NVE ensemble, for larger samples, and using Langevin dynamics (results not shown).
{\bf Exp:} 500 particles interacting via a pair potential with exponential repulsion; $U(r) = \frac{\epsilon}{8} \left[6e^{-14(r/\sigma - 1)}-14(\sigma/r)^6\right]$, simulated with Metropolis dynamics.
{\bf Kob-Andersen binary LJ:} The Kob-Andersen binary Lennard-Jones liquid, N=1000 \cite{kob94}. This includes data for the less-viscous liquid, the highly viscous liquid, as well as the glass. 
{\bf Asymmetric dumb-bell:} 512 asymmetric ``dumb-bell'' molecules \cite{ped06}.
{\bf OPLS-UA Toluene:} A 7-site united-atom model of toluene, N=1000 \cite{jor84}.
{\bf GROMOS Methanol:} 512 methanol molecules \cite{sco99}.
{\bf SPC/E water:} 4142 SPC/E water molecules \cite{ber87}.
Except for the two systems with Coulomb interactions (Methanol and water), all systems studied have strong correlations between fluctuations in virial and potential energy; correlation coefficients are above 0.9 for all state-points shown, except those with negative pressure, where they are slightly smaller. The correlation coefficients increase with increasing density and temperature.}
\label{figure_4}
\end{center}
\end{figure}

In order to investigate how general the $W,U$ correlations are, several other systems were studied. If $W(t)$ and $U(t)$ are perfectly correlated ($R=1$), the following identity applies: $\langle\Delta W \Delta U\rangle^2=\langle(\Delta W)^2\rangle\langle(\Delta U)^2\rangle$. Fig. 2 summarizes our simulations in a plot where the diagonal corresponds to perfect correlation and the y-variable by the fluctuation-dissipation theorem equals $T$ times the configurational pressure coefficient [$(T/V)(\partial W/\partial T)_V$]. Liquids with strong $W,U$ correlations ($R>0.9$) include: 1) A liquid with exponential short-range repulsion; 2) The Kob-Andersen binary Lennard-Jones liquid \cite{kob94}; 3) A liquid consisting of  asymmetric ``dumb-bell'' type molecules (two unlike Lennard-Jones spheres connected by a rigid bond \cite{ped06}); 4) A 7-site united-atom model of toluene \cite{jor84}.  The last three liquids are examples of good glass-formers that can be cooled to high viscosity without crystallizing. Liquids not showing strong $W,U$ correlations are methanol \cite{sco99} and SPC/E water \cite{ber87}; in these models the instantaneous potential energy has contributions from both LJ interactions ($U_{LJ}(t)$) and Coulomb interactions ($U_{C}(t)$). Since the Coulomb interaction is an inverse power-law with $n=1$, the corresponding contribution to the instantaneous virial is given by $W_{C}(t)=U_{C}(t)/3$, i.e., perfect correlation. For the LJ interaction of SPC/E water we find $\Delta W_{LJ}(t)\approx 6\Delta U_{LJ}(t)$ with correlations coefficients above 0.9. Since the proportionality constants are different, however, the \emph{sums} of the contributions do not correlate very well. In fact, close to the density maximum of water we find that W(t) and U(t) are uncorrelated. For methanol  $\Delta W(t)$  and $\Delta U(t)$ correlate well at such high temperatures that the LJ interactions completely dominate ($\approx$3000K).

\begin{figure}
\begin{center}
\includegraphics[width=8cm]{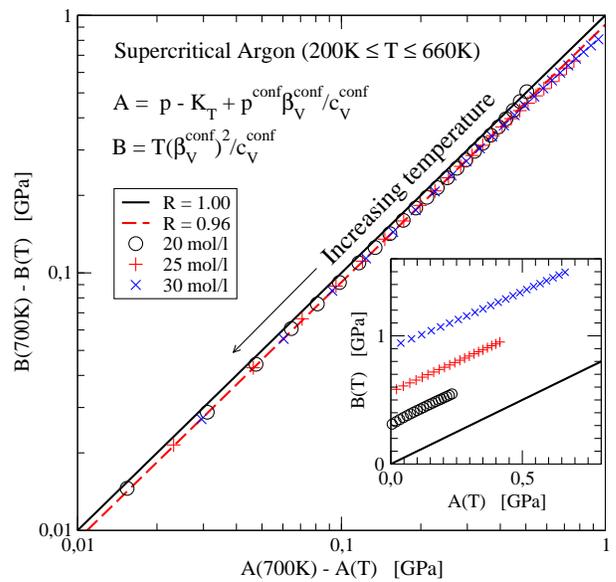}
\caption{
Data for supercritical Argon at 3 different densities covering the temperature range 200K-660K \cite{lin05} showing a $W,U$ correlation of 96\% [$K_T\equiv -V(\partial p/\partial V)_T$, $p^{\rm conf}\equiv p-Nk_BT/V = W/V$, $\beta_V^{\rm conf}\equiv(\partial p/\partial T)_V-Nk_B/V$, and $c_V^{\rm conf}\equiv C_V/V-(3/2)Nk_B/V$]. The full line corresponds to perfect correlation between virial and potential energy \cite{Suplement}. The inset shows the prediction for an exact inverse power-law interatomic potential (full line): $B(T) = A(T)$. The poor fit shows that the fact that {\it fluctuations} are well described by an effective inverse power-law does not imply that this is the case also for the equation of state.}
\label{figure_5}
\end{center}
\end{figure}

Do strong pressure-energy correlations have consequences accessible by experiment? In the following we demonstrate how it is possible to test for strong W,U correlations in systems where the kinetic contribution to the isochoric heat capacity is known, exemplified by experimental data for supercritical argon. From the definition of $R$, utilizing three fluctuation formulas \cite{all93} it is straightforward to show \cite{Suplement} (where $c_V^{\rm conf}$ and $\beta_V^{\rm conf}$ are the configurational parts of the isochoric heat capacity per volume and pressure coefficient respectively, and $K_T$ is the isothermal bulk modulus) that:

\begin{equation}
  T{\left( \beta_V^{\rm conf} \right)^2}/{c_V^{\rm conf}} = R^2\left( p - K_T + {\langle X \rangle}/{V}  \right) \label{PowerPred}
\end{equation}
Here $X= \sum_{i<j} r w'(r_{ij})/9$ is the so-called ``hypervirial'' where $w(r)=r \phi'(r)$. This quantity cannot be determined experimentally \cite{all93}, so we apply an approximation. For an exact power-law potential, one has $X = (n/3) W$, which, however, is not expected to be a good approximation since the apparent power-law depends on the state-point. In the vicinity of a given reference state-point, however, one expects $[X-X_{ref}]\approx (n/3) [W-W_{ref}]$ along an isochore (confirmed by our simulations). Using this approximation and assuming that $R$ is roughly constant, we can test for strong W,U correlations. Fig. 3 shows experimental data for supercritical Argon covering the temperature range 200K-660K at three different densities \cite{lin05}, showing that W and U correlate  96\% in this case. The apparent power-law exponents, $n=3\beta_V^{\rm conf}/c_V^{\rm conf}$ \cite{Suplement} varies from 13.2 to 15.8, decreasing with increasing temperature and density. The inset shows that the Argon data do \emph{not} follow the prediction following from an exact inverse power-law potential \cite{Suplement,pow} (Eq.{}(\ref{PowerPred}) with $R=1$ and $X = (n/3) W$); thus the W,U correlations show that such an effective power-law description applies to a good approximation only for the {\it fluctuations}. This is analogous to the situation for the Lennard-Jones liquid: The equation of state is poorly described by that following from an inverse power-law potential (see e.g. \cite{joh93}), although the fluctuations are well described by this.

A different class of systems where it is possible to test for strong W,U correlations experimentally is highly viscous liquids \cite{ped06,ell07}. These are characterized by a clear separation of time scales between the fast vibrational degrees of freedom on the picosecond time scale and the much slower configurational degrees of freedom on the second or hour time scale, depending on temperature \cite{kau48,bra85,ang00,deb01,bin05,sci05,dyr06}. Suppose a highly viscous liquid has perfectly correlated $W,U$ fluctuations. When $W$ and $U$ are time-averaged over, say, one tenth of the liquid relaxation time \cite{ped06}, they still correlate 100\%. Since the kinetic contribution to pressure is fast, the time-averaged pressure equals the time-average of $W/V$ plus a constant. Similarly, the time-averaged energy equals the time-averaged potential energy plus a constant. Thus the fluctuations of the time-averaged $W$ and $U$ are the slowly fluctuating parts of pressure and energy, so these slow parts will also correlate 100\% in their fluctuations. This is the single ``order'' parameter scenario of Ref. \cite{ell07}. In this case, knowledge of just one of the eight fundamental frequency-dependent thermoviscoelastic response functions implies knowledge of them all \cite{ell07} (except for additive constants \cite{note2}). This constitutes a considerably simplification of the physics of glass-forming liquids. Unfortunately, there are few reliable data for the frequency-dependent thermoviscoelastic response functions \cite{chr07}. Based on the results presented above we predict the existence of a class of ``strongly correlating liquids'' where just one frequency-dependent thermoviscoelastic response function basically determines all. 
Our simulations suggest that the class of strongly correlating liquids includes van der Waals liquids, but not network liquids like water or silica. This is consistent with the findings of De Michele et. al. \cite{dem06}. 

Very recently Coslovich and Roland studied diffusion constants $D$ in highly viscous binary Lennard-Jones mixtures at varying pressure and temperature \cite{cos07}. Their data follow the ``density scaling'' expression $D=F(\rho^\gamma/T)$ \cite{scaling}, and they showed convincingly that the exponent $\gamma$ reflects the effective inverse power law of the repulsive core. In view of these findings, we conjecture that  {\it strongly correlating viscous liquids obey density scaling, and vice versa}. If this conjecture is confirmed, by virtue of their simplicity the class of strongly correlating liquids provides an obvious starting point for future theoretical works on the highly viscous liquid state.

\acknowledgments 
{\bf Acknowledgments:} The authors wish to thank S{\o}ren Toxv{\ae}rd for useful discussions. 
This work was supported by the Danish National Research Foundation's (DNRF) centre for viscous liquid dynamics ``Glass and Time.''


\begin{thebibliography}{99}

\bibitem{lan70} L. D. Landau and E. M. Lifshitz,  {\it Statistical Physics} Part 1 (Pergamon Press, London, 1980).

\bibitem{han86} J. P. Hansen and I. R.  McDonald, {\it Theory of Simple Liquids}, 2nd ed. (Academic Press, New York, 1986).

\bibitem{rei98} L. E. Reichl, {\it A Modern Course in Statistical Physics}, 2nd ed. (Wiley, New York, 1998).

\bibitem{all93} M. P. Allen and D. J. Tildesley, {\it Computer Simulation of Liquids} (Oxford Science Publications, Oxford, 1987).

\bibitem{gromacs} H. J. C. Berendsen, D. van der Spoel, and R. van Drunen, Comp. Phys. Comm. {\bf 91}, 43 (1995); 
E. Lindahl, B. Hess, and D. van der Spoel,  J. Mol. Mod. {\bf 7}, 306 (2001).

\bibitem{len31} J. E. Lennard-Jones,  Proc. Phys. Soc. London {\bf 43}, 461 (1931).

\bibitem{NoseHoover} S. A. Nos\'e,  J. Chem. Phys. {\bf 81}, 511 (1984); 
W. G. Hoover,  { Phys. Rev. A} {\bf 31}, 1695 (1985).

\bibitem{kan85} H. S. Kang, C. S. Lee, T. Ree, and F. H. Ree, J. Chem. Phys. {\bf 82}, 414 (1985).

\bibitem{BenAmotz03} D. Ben-Amotz and G. J. Stell, J. Chem. Phys. {\bf 119}, 10777 (2003).

\bibitem{wee71} J. D. Weeks, D. Chandler, and H. C. Andersen, J. Chem. Phys. {\bf 54},  5237 (1971).

\bibitem{kob94} W. Kob and H. C. Andersen,  Phys. Rev. Lett. {\bf 73}, 1376 (1994).

\bibitem{ped06} U. R. Pedersen, T. Christensen, T. B. Schr{\o}der, and J. C. Dyre,  cond-mat/0611514.

\bibitem{jor84} W. L. Jorgensen, J. D. Madura, and C. J. Swenson, J. Am. Chem. Soc. {\bf 106}, 6638 (1984).

\bibitem{sco99} W. R. P. Scott, P. H. Hunenberger, I. G.  Tironi, {et al.}, J. Phys. Chem. A {\bf 103}, 3596 (1999). 

\bibitem{ber87} H. J. C. Berendsen, J. R.  Grigera, and T. P.  Straatsma, J. Phys. Chem. {\bf 91}, 6269 (1987).

\bibitem{Suplement} See the supplementary material.

\bibitem{lin05} E. W. Lemmon, M. O. McLinden and D. G. Friend, ``Thermophysical Properties of Fluid Systems'' in
{\it NIST Chemistry WebBook, NIST Standard Reference Database Number 69}, Eds. P.J. Linstrom and W.G. Mallard, June 2005,
National Institute of Standards and Technology, Gaithersburg MD, 20899 (http://webbook.nist.gov).

	\bibitem{pow} J. N. Cape and L. V. Woodcock, J. Chem. Phys. {\bf 72} 976 (1980);
	M. Baus and J.-P. Hansen, Phys. Rep. {\bf 59}, 1 (1980);  
	J. D. Weeks, Phys. Rev. B {\bf 24}, 1530 (1981). 

\bibitem{joh93}
J. K. Johnson, J. A. Zollweg, and K. E. Gubbins, Mol. Phys. {\bf 73}, 591 (1993).

\bibitem{ell07} 
N. L. Ellegaard, T. Christensen, P. V. Christiansen, N. B. Olsen, U. R. Pedersen, T. B. Schr{\o}der, and J. C. Dyre, J. Chem. Phys. {\bf 126}, 074502 (2007).

\bibitem{kau48} W. Kauzmann, Chem. Rev. {\bf 43}, 219 (1948).

\bibitem{bra85} S. Brawer, {\it Relaxation in viscous liquids and glasses} (American Ceramic Society, Columbus, OH, 1985).

\bibitem{ang00} C. A. Angell, K. L. Ngai, G. B. McKenna, P. F. McMillan, and S. W. Martin, J. Appl. Phys. {\bf 88}, 3113 (2000).

\bibitem{deb01} P.G. Debenedetti and F. H.  Stillinger,  Nature {\bf 410}, 259 (2001).

\bibitem{sci05} F. Sciortino, J. Stat. Mech., P05015 (2005).

\bibitem{bin05} K. Binder and W. Kob, {\it Glassy Materials and Disordered Solids: An Introduction to their Statistical Mechanics} (World Scientific, Singapore, 2005).

\bibitem{dyr06} J. C. Dyre,  Rev. Mod. Phys. {\bf 78}, 953 (2006).

\bibitem{note2} More precisely, knowledge of one response function implies knowledge of any other except for two real numbers giving the low- and high-frequency limits, respectively.

\bibitem{chr07} T. Christensen, N. B. Olsen, and J. C. Dyre, 
Rev. E {\bf 75}, 041502 (2007). 

\bibitem{dem06} C. De Michele, P.  Tartaglia, and F. Sciortino, J. Chem. Phys. {\bf 125}, 204710 (2006).

\bibitem{cos07} D. Coslovich and C. M. Roland, arXiv:0709.1090 (2007).

\bibitem{scaling} C. Alba-Simionesco, A. Cailliaux, A. Alegria, and G. Tarjus, Europhys. Lett. {\bf 68}, 58 (2004); 
R. Casalini and C. M. Roland, Phys. Rev. E {\bf 69}, 062501 (2004); 
C. M. Roland, S. Hensel-Bielowka, M. Paluch, and R. Casalini, Rep. Prog. Phys. {\bf 68}, 1405 (2005).

\end{thebibliography}
\end{document}